\documentclass[aps,prl,twocolumn,superscriptaddress,showpacs]{revtex4}
\usepackage{graphicx}
\usepackage{bm}
\usepackage{color}
\input colordvi
\begin{document}
%
%
%
\title{Quantum Dynamics of a Driven Correlated System Coupled to Phonons}
%
\author{L. Vidmar}
\affiliation{J. Stefan Institute, SI-1000 Ljubljana, Slovenia }
\author{J. Bon\v{c}a}
\affiliation{Department of Physics, FMF, University of Ljubljana, Jadranska 19, SI-1000 Ljubljana, Slovenia }
\affiliation{J. Stefan Institute, SI-1000 Ljubljana, Slovenia }
\author{T. Tohyama}
\affiliation{Yukawa Institute for Theoretical Physics, Kyoto University, Kyoto 606-8502, Japan }
\author{S. Maekawa}
\affiliation{The Advanced Science Research Center, Japan Atomic Energy Agency, Tokai 319-1195, Japan }
\affiliation{CREST, Japan Science and Technology Agency, Sanbancho, Tokyo 102-0075, Japan }
\begin{abstract}
Nonequilibrium interplay between charge, spin and lattice degrees of freedom on a square lattice is studied for  a single charge carrier doped in  the $t$--$J$--Holstein model.
In the presence of a static electric field we calculate the quasistationary state.
With increasing electron-phonon (e-ph) coupling the carrier mobility decreases, however, we find increased steady state current due to e-ph coupling in the regime of negative differential resistance.
We explore the distribution of absorbed energy between the spin and the phonon subsystem.  For model parameters as relevant for cuprates, the majority of the gained energy flows into the spin subsystem.
\end{abstract}

\maketitle

\textit{Introduction.}---One of the outstanding  contemporary challenges in condensed matter physics is to understand dynamics of interacting quantum systems exposed to an external perturbation. Advanced pump and probe techniques with few femtosecond time-resolution and broad-band THz spectroscopy were developed~\cite{wall,pashkin,gadermaier,kawakami} to drive the system out of equilibrium and measure its nonequilibrium physical  properties. These measurements were  complemented by time-resolved photoemission spectroscopy~\cite{perfetti07,cortes}, electron cristallography~\cite{t-cristall} and Raman scattering~\cite{t-Raman}.
In the systems with competing interactions the most demanding task is to disentangle different elementary excitations arising at comparable energy-time scales.
Possibly the most studied example of such materials are cuprate superconductors where 
the influence of the  electron-phonon (e-ph) interaction~\cite{alex} on ultrafast dynamics was recently investigated, and different mechanisms were proposed~\cite{gadermaier,perfetti07,
t-Raman,okamoto,pheno,t-BCS,nontherm}.
Considerable effort has been devoted as well to study response of pump-excited Mott insulators~\cite{trpes} and to unravel  
complex thermalization processes of strongly correlated systems~\cite{eckstein11}.

Despite a considerable ongoing effort to understand nonequilibrium dynamics of quantum systems, a sizable gap perseveres between theory and experiments.
A theoretical insight into many-body quantum phenomena far from equilibrium has been obtained, among others, through works on nonlinear transport of half-filled Hubbard systems~\cite{breakdown,breakdown_dmft,aron,amaricci}
and Falicov--Kimball model~\cite{bloch}.
It was found that to obtain a non-zero steady current in the metallic Hubbard model driven by a static electric field, it is crucial to couple the system to a heat reservoir~\cite{aron,amaricci}, which prevents an uncontrollable heating of the system.
In contrast, the aim of our paper is to study a doped strongly correlated system coupled to phonons, where the energy gained by the motion of a charge carrier along 
the field is absorbed by quantum spin and phonon degrees of freedom which are all explicitly included in the model, 
and the influence of environment is supposed to show up on a much longer timescales.
We study a single charge carrier doped into two-dimensional (2D) plane within the $t$--$J$--Holstein model, which is a prototype model for the description of competing interactions in cuprates.
Hereby, we address a fundamental,  yet unresolved  question concerning  the interplay between strong correlations and e-ph interaction in a driven quantum system far from equilibrium.

So far, nonequilibrium response of the generalized Hubbard--Holstein model has been analyzed on a 1D chain~\cite{matsueda} and on 8-site 2D cluster~\cite{kawakami}.
While qualitatively different behavior is expected in 1D systems due to a spin-charge separation~\cite{1d}, a detailed investigation of 2D systems is still pending.
By generalizing recently developed method \cite{holE,tjE}, we drive the system by a static electric field until it reaches a quasistationary (QS)  state  with  a constant current and  a constant  energy flow to the system. To our knowledge this is the  first study of a  2D strongly correlated system (SCS) coupled to phonons where QS conditions are achieved.

In this Letter, we explore two important aspects of nonequilibrium carrier dynamics:
\textit{(i)} We establish the influence of e-ph coupling  on the nonlinear transport properties of a carrier in SCS.
We show that the coupling to phonons decreases carrier mobility, however, it leads  to an enhancement of QS current in the regime of negative differential resistance;
\textit{(ii)} We compare  the energy absorbed by the spin subsystem and the one  absorbed by lattice vibrations. 
Taking into account model parameters fitting  cuprates we find that the spin subsystem absorbs the energy from the field more efficiently than the lattice.

\textit{Model and numerical method}---We define the time--dependent $t$--$J$--Holstein Hamiltonian as
\begin{eqnarray}
\vspace*{-0.0cm}
H &=& -t_0 \sum_{\langle \mathbf{l}\mathbf{j} \rangle,\sigma} \left[ {\mathrm e}^{i \phi_{\mathbf{l}\mathbf{j}}(t)}\; \tilde{c}^{\dagger}_{\mathbf{l}, \sigma}
\tilde{c}_{\mathbf{j}, \sigma} +{\mathrm H.c.} \right]+J \sum_{\langle \mathbf{l}\mathbf{j} \rangle} \mathbf{S_l} \cdot \mathbf{S_j}, \nonumber \\
 &+& {g} \sum_{\mathbf{j}} n_{\mathbf{j}} (a_{\mathbf{j}}^\dagger + a_{\mathbf{j}}) +
\omega_0\sum_{\mathbf{j}} a_{\mathbf{j}}^\dagger  a_{\mathbf{j}}, \label{ham}
\end{eqnarray}
where $\tilde{c}_{\mathbf{j},\sigma}=c_{\mathbf{j},\sigma}(1-n_{\mathbf{j},-\sigma})$ is a projected fermion operator, $\phi_{\mathbf{l}\mathbf{j}}(t)$ is time--dependent magnetic flux and $\langle \mathbf{l}\mathbf{j} \rangle$ denote nearest neighbors.
The charge carrier is coupled to Einstein phonons with  the energy $\omega_0$ via e-ph coupling constant $g$, 
where $a_{\mathbf{j}}$ is the  phonon annihilation operator and $n_{\mathbf{j}} = \sum_{\sigma} n_{\mathbf{j,\sigma}}$.
The ground state at $\mathbf{k}_0=(\pi/2,\pi/2)$ is calculated by exact diagonalization defined over a limited functional space (EDLFS)~\cite{edlfs,pol,bipol}.
To construct  functions of the Hilbert space we use the off-diagonal parts of Eq.~(\ref{ham}) in the basis generator 
$\left\{|\varphi_{l}^{N_h} \rangle \right\}=
[H_{t_0}(\phi_{\mathbf{l}\mathbf{j}}=0) + \tilde{H}_J + H_{\rm g}]^{N_h} |\varphi_0 \rangle,$ 
where $|\varphi_0\rangle=c_{\mathbf{k}_0}|\mbox{N\' eel}\rangle$ represents a translationally invariant state of a carrier in the N\' eel background.
We switch on the static uniform electric field $F$ along the diagonal at time $t=0$ and perform the time evolution by iterative Lanczos method~\cite{lantime}.
Accordingly, we define the charge current along the diagonal $j(t)$ and set ${\phi}_{\mathbf{l}\mathbf{j}}(t)=-Ft/\sqrt{2}$ for (positive) $\hat{x}$-- and $\hat{y}$--direction.
We measure $F$ in units of $[ t_0/e_0 a ]$ and set $t_0=e_0=a=1$.
Recently, the EDLFS method was reported to effectively  calculate the QS state of a doped charge carrier within the 2D $t$--$J$ model~\cite{tjE} as well as of the Holstein polaron~\cite{holE}.
The strength of the numerical method is in construction of  the Hilbert space that enables  not only an accurate description  of the ground state of the spin--lattice polaron, but it allows for enough extra spin and phonon excitations to absorb energy, emitted by the field driven carrier, until the system reaches the QS state.

\begin{figure}
\includegraphics[width=0.47\textwidth]{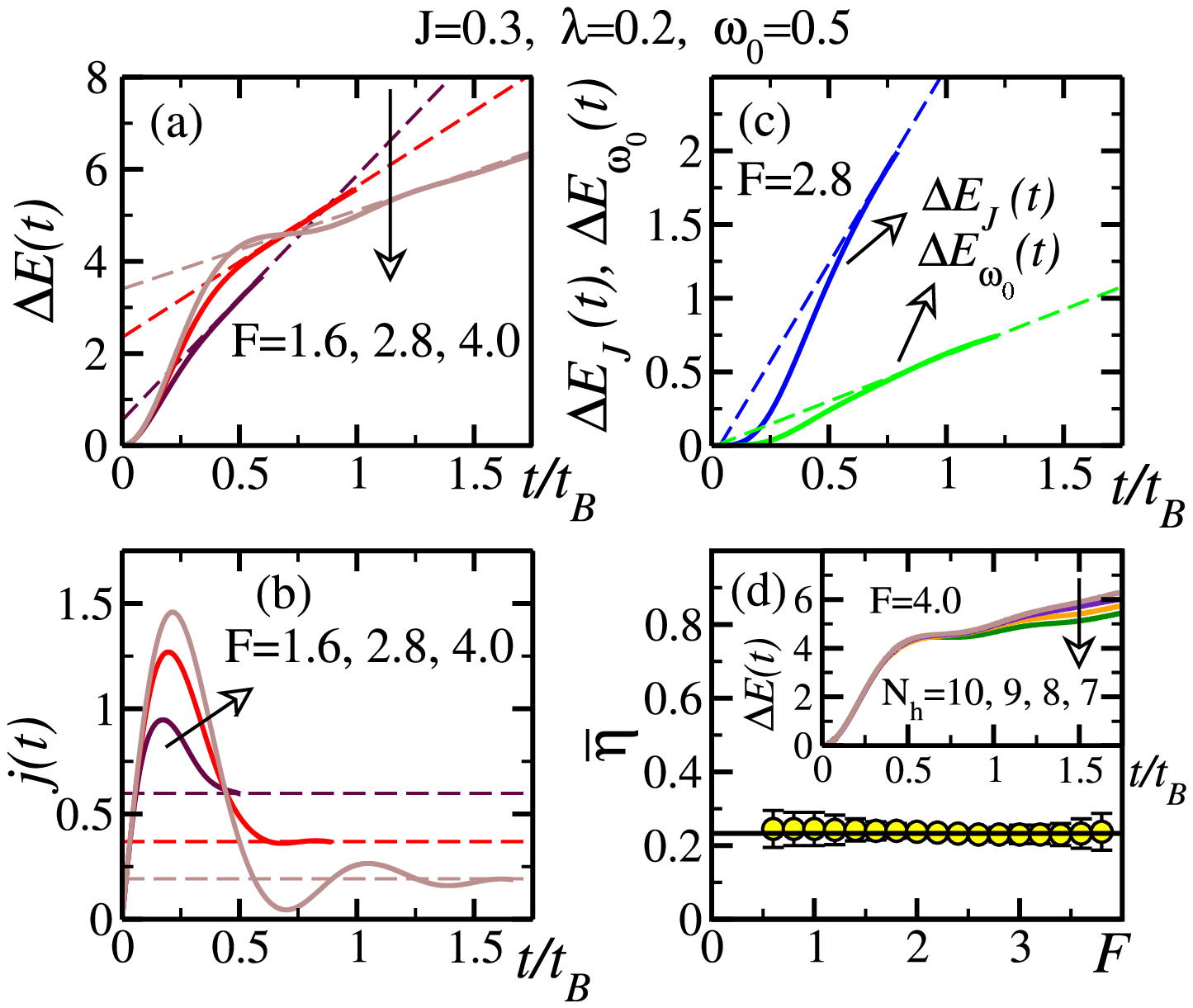}
\caption{(Color online) 
(a) $\Delta E(t)$ and (b) $j(t)$ for $F=1.6, 2.8$ and $4.0$.
We measure time in units of Bloch time $t_B$, where $\omega_B=F/\sqrt{2}$.
Dashed lines in (a) represent extrapolation of linear increase of $\Delta E(t)$ in the QS state and dashed lines in (b) the corresponding QS current $\bar{j}$.
We set $J=0.3$, $\lambda=0.2$ and $\omega_0=0.5$.
Numerical accuracy of the time propagation is checked by the total-energy-gain sum rule $\Delta E(t) = F \int_0^t j(t')dt'$.
(c) $\Delta E_J(t)$ and $\Delta E_{\omega_0}(t)$ for $F=2.8$. Dashed lines represent their extrapolation to the QS state.
(d) Energy distribution ratio in the QS state $\bar \eta$ vs. F. 
Inset: Convergence of $\Delta E(t)$ for $N_h=7,8,9,10$ with $N_{st}=3.7 \times 10^4, 1.6 \times 10^5, 6.7 \times 10^5$ and $2.9 \times 10^6$, respectively.
}
\label{fig1}
\end{figure}

\textit{Results.}---We focus mostly on weak and moderate values of e-ph coupling $\lambda=g^2/8t_0\omega_0$ and different regimes of $\omega_0$, while keeping $J=0.3$ constant.
Results describing  the real-time propagation of a spin-lattice polaron are shown in Fig.\ref{fig1}(a)-(c) for $\lambda=0.2$ and $\omega_0=0.5$.
After a short transient regime $t/t_B\lesssim 1$ the system enters the QS state with the steady   current $j(t)=\bar j$  and the linear increase of the total energy, satisfying $\Delta\dot{E}(t)=F \bar{j}$ (compare Fig.~\ref{fig1}(a) and (b)),
where $\Delta E(t)= \langle H(t)\rangle-\langle H(0)\rangle$. 
The longest  time of propagation in the QS regime  is limited due to a finite number  of spin and phonon excitations in the  Hilbert space acting as  reservoirs  for the energy absorption.  They are determined roughly by $N_h$, and choosing $N_h=10$ enables computation of QS quantities with high accuracy
(see the inset of Fig.~\ref{fig1}(d) for comparison with lower $N_h$).
In this work, we are interested in values of $F$ when the response of the system is dissipative, {\it i.e.}, $\bar j \neq 0$.
The time evolution for $F\ll 1$ becomes  adiabatic with $\bar j = 0$, as pointed out in discussion of Ref.~\cite{tjE}.
In the opposite limit $F\gg 1$, $F$ should not exceed the threshold electric field $F_{th}$ required for the breakdown of Mott insulator at half-filling.
Estimating $F_{th}$ from the recent DMFT study ($F_{th}\sim 10$ in our units, $J=0.3$)~\cite{breakdown_dmft}, the values of $F$ indeed fulfill this condition.

We next turn to the calculation of energy flow to the spin and phonon subsystem.
In contrast to the nonequilibrium studies of closed systems at half-filling where the current response is considerably influenced by the Joule heating~\cite{tv},
the problem of a single carrier in dissipative medium enables investigation of  the steady growth of energy due to carrier propagation in initially undistorted background at $T=0$.
In Fig.\ref{fig1}(c) we show $\Delta E_J(t)$ and $\Delta E_{\omega_0}(t)$, {\it i.e.}, expectation values of the second and fourth term of Eq.~(\ref{ham}).
The  energy flows to  both subsystems are determined by ${\cal P}_{J,\omega_0}(t)=\Delta \dot{E}_{J,\omega_0}(t)$,  defining the distribution ratio $\eta(t)={\cal P}_{\omega_0}(t)/{\cal P}_J(t)$.
In the QS state both, $\Delta E_J(t)$ and $\Delta E_{\omega_0}(t)$, reveal a linear time dependence (dashed lines in  Fig.~\ref{fig1}(c)) and therefore $\eta(t)=\bar \eta$.

In Fig.~\ref{fig1}(d) we show $\bar \eta (F)$ which exhibits only tiny variation around the constant value.
This result is rather surprising since $F$ strongly  influences  the  energy flow  into the system, nevertheless,  $\bar \eta$ remains fairly field-independent.

\begin{figure}
\includegraphics[width=0.47\textwidth]{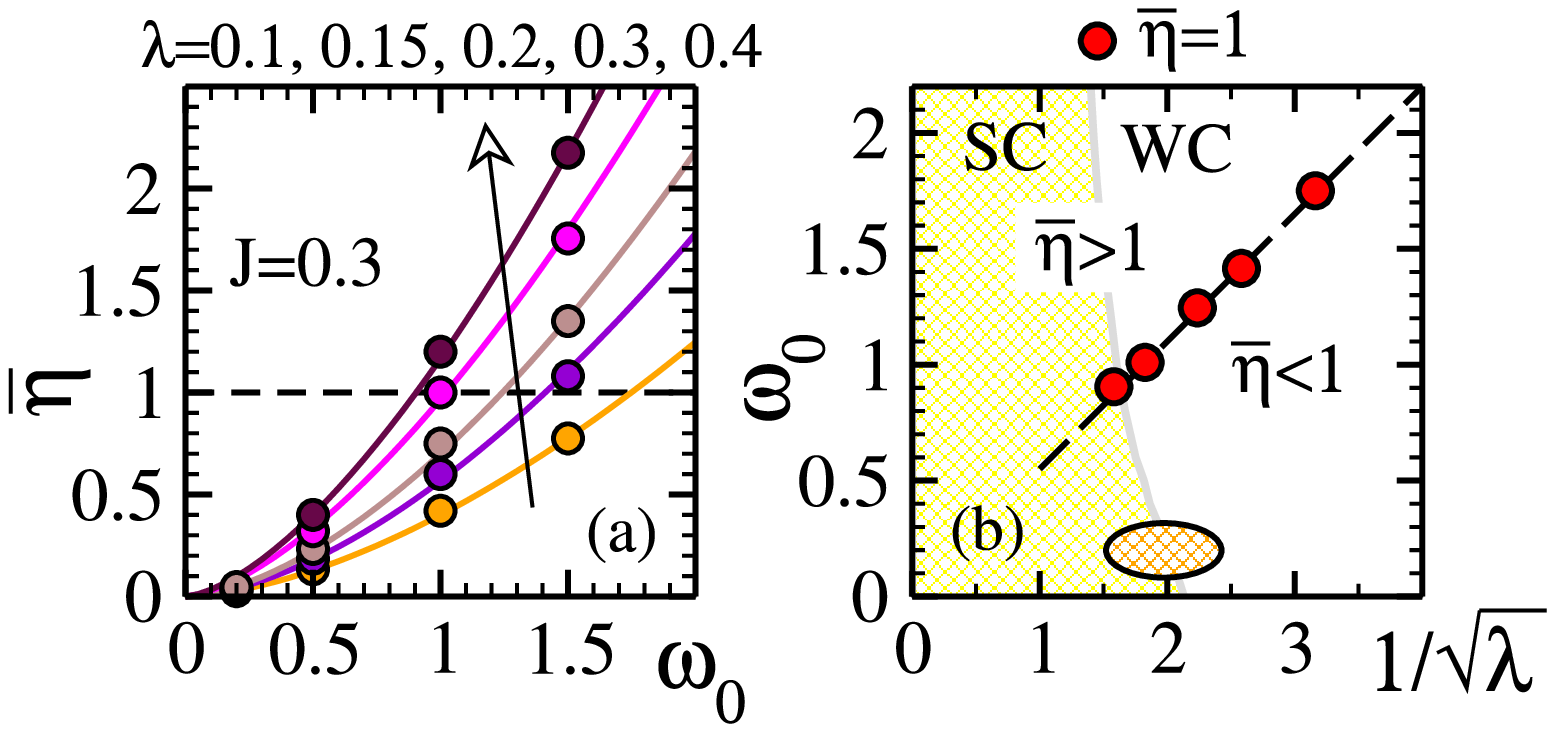}
\caption{(Color online) 
(a) $\bar \eta$ vs. $\omega_0$ for different $\lambda$ and $J=0.3$.
At $\omega_0=0.2$, $\bar \eta$ is calculated for $\lambda \leq 0.2$, while for higher $\omega_0$ a $N_h$-independent $\bar \eta$ is obtained until $\lambda=0.4$.
(b) Red dots denote values of $\omega_{0,\bar \eta=1}$ for which the condition $\bar \eta=1$ is fulfilled, and dashed line represents a fit $\omega_{0,\bar \eta=1}=c/\sqrt{\lambda}$, with $c=1.83$.
Left and right side of the plot correspond to the strong coupling (SC) and weak coupling (WC) regime of e-ph coupling, respectively.
WC-SC crossover was calculated by the EDLFS method~\cite{pol} and is in agreement with other numerical methods~\cite{mishchenko04,prelovsek06}.
Filled ellipse represents parameters as relevant for cuprates.
}
\label{fig2}
\end{figure}


This result facilitates the investigation   of the efficiency of the energy absorption through spin and phonon channel when e-ph coupling and phonon frequencies are varied.
With increasing $\omega_0$, $\bar \eta$ increases at fixed $\lambda$ as seen in Fig.~\ref{fig2}(a).
We are particularly interested in the case $\bar \eta=1$, {\it i.e.}, when a propagating carrier deposits equal amount of the gained energy to the spin and the phonon subsystem alike.
For this purpose we calculate $\omega_{0,\bar \eta=1}$ 
and plot it vs. $\lambda$ in Fig.~\ref{fig2}(b).
Remarkably, a scaling $\omega_{0,\bar \eta=1}\sim 1/\sqrt{\lambda}$ is found. In the weak coupling regime the energy flow to phonons dominates over the flow to the spin subsystem only for large values of $\omega_0>1$. 
Recent photo-emission~\cite{mishchenko04,pol} and optical experiments~\cite{optic,opt} on cuprates that were interpreted within the $t$-$J$-Holstein model, assigned the realistic $\lambda$ to be in the interval $[0.2,0.3]$ and $\omega_0\sim 0.2$, 
as indicated by the filled ellipse in Fig.~\ref{fig2}(b). In the parameter regime  as relevant for cuprates, the majority of the absorbed energy via the charge carrier driven by the constant electric field  flows  in  the spin subsystem.

\begin{figure}
\includegraphics[width=0.47\textwidth]{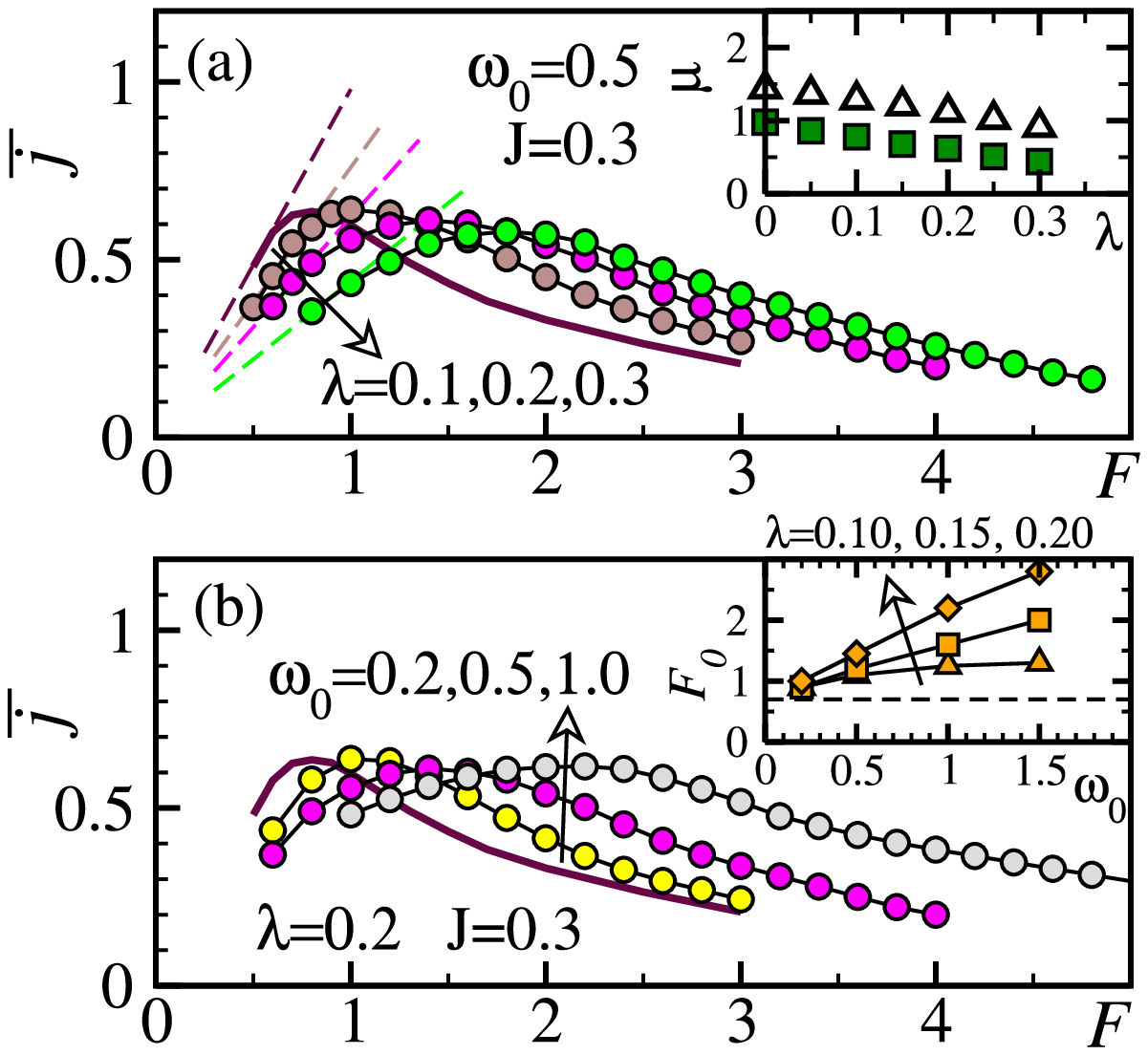}
\caption{(Color online)
$\bar{j}$--$F$ characteristics of the QS state.
(a) $\bar{j}$--$F$ at $\omega_0=0.5$ and (b) at $\lambda=0.2$.
Solid line shows $\bar{j}$--$F$ at $\lambda=0$.
Inset of (a): $\mu=\bar{j}/F$ (squares) calculated from the dashed lines in (a), LR mobility $\mu_{LR}=2\pi D$ (triangles) with charge stiffness $D$ as calculated in Ref.~\cite{opt}.
Maximal current shows slight dependence on $N_h$ (compare $\bar{j}$--$F$ at $\lambda=0$ with Ref.~\cite{tjE}), and contributes to discrepancy between $\mu$ and $\mu_{LR}$.
Inset of (b): crossover field $F_0$ between PDR and NDR regime vs. $\omega_0$ for different $\lambda$. Dashed line shows $F_0$ for $\lambda=0$.
}
\label{fig3}
\end{figure}

We next explore the influence of e-ph coupling on carrier's QS propagation. 
In Fig.~\ref{fig3} we plot $\bar{j}$--$F$ characteristics calculated in the QS state.
Solid line represents $\bar{j}$--$F$ characteristics for a plain $t$-$J$ model at $J=0.3$, as calculated in Ref.~\cite{tjE}.
In the latter work, it was shown that a regime of positive differential resistivity (PDR) at small $F$ evolves into a negative differential resistivity (NDR) regime at crossover field $F_0\sim 2.3 J$.
The effect of increasing $\lambda$, displayed in Fig.~\ref{fig3}(a) at $\omega_0=0.5$, is to extend the region of PDR toward larger $F$, and to decrease the carrier mobility $\mu$ as shown in the inset.
A similar  tendency is observed when increasing  $\omega_0$  while keeping $\lambda$ fixed, see Fig.~\ref{fig3}(b).
The carrier mobility in the PDR regime calculated from the $\bar{j}$--$F$ characteristics 
is in qualitative agreement with the linear response (LR) theory~\cite{opt}, and allows one to detect the limit of LR regime occurring below the crossover field $F_0$.
The variation of $F_0$ with $\lambda$ and $\omega_0$ is shown in the inset of Fig.~\ref{fig3}(b).

While a decrease of mobility with larger e-ph coupling in the PDR regime can be intuitively understood due to increased scattering on phonon excitations, the most intriguing result of Fig.~\ref{fig3} represents the phonon-induced enhancement of $\bar{j}$ in the NDR regime.
In general, appearance of NDR regime for large $F$ is a consequence of limited degrees of freedom contained  in the model that are available to  absorb the excess energy, which impedes the carrier motion along the field.
For instance, a NDR regime in 2D $t$-$J$ model is characterized by pronounced  transverse oscillations of the carrier that  serve to emit the excess energy (gained by hopping along the filed direction) to spin excitations~\cite{tjE}.
Alternatively, carrier propagation can be for large $F$ interpreted on the basis of Wannier-Stark (WS) states where phonon  assisted hopping between these states leads to nonzero $\bar{j}$~\cite{emin,holE}.
In this picture one can explain two phenomena observed for large $F$ in Fig.~\ref{fig3}(a): 
$(i)$ with increasing $F$, the overlap between neighboring WS states mediated by  the combined phonon- and magnon- assisted hopping decreases, hence $\bar{j}$ decreases with $F$,
$(ii)$ increasing  the e-ph interaction  seems to boost the already existing magnon- mediated overlap between WS states at $\lambda = 0$. In contrast to low $F$, where additional scattering on phonons at $\lambda > 0$ diminishes the carrier mobility, at large $F$ the increase of  $\bar j$ is due to  opening of additional channels for depositing the excess energy through simultaneous   emission of magnons as well as phonons. 
This explains the seemingly counterintuitive effect observed in our results, i.e., that the current is \textit{enhanced} due to an increased e-ph interaction.
In a similar fashion, increasing of $\omega_0$ enhances  the phonon assisted hopping between WS states at fixed $\lambda$ and leads to an enhancement of $\bar{j}$ shown in Fig.~\ref{fig3}(b).

\textit{Discussion and Conclusion.}---By investigating the nonequilibrium charge dynamics in strongly correlated medium coupled to phonons,
we showed that the ratio of the energy flow to the phonon relative to the spin subsystem remains field-independent.
Therefore, even though the values of $F$ considered here are rather large, we expect that the calculated distribution of the energy flow holds generally for any values of $F$ for a weakly doped system where the dominant interactions are described by Eq.~\ref{ham}.
In particular, for  model parameters relevant for  cuprates,  the energy flow into the spin subsystem remains dominant.  This result may signal stronger coupling  of charge carrier to the spin system in comparison to e-ph coupling.
An intuitive physical picture emerges when considering hopping of the hole in the spin background, simultaneously coupled to phonons. The hole can hop a few lattice spacings without exciting a single phonon, however each hop of the hole through a N\' eel background unavoidably  generates spin excitations. 

The influence of phonons on carrier's QS propagation shows strong dependence on the particular regime of the $\bar{j}$--$F$ characteristics.
While increasing e-ph coupling  leads to a decrease of the carrier mobility at small $F$, the QS current increases due to increased coupling to phonons for large $F$, where the system enters NDR.
Even though the calculation of a stable QS current in recent theoretical studies of SCS is possible only by coupling the system to the heat reservoir~\cite{aron,amaricci,tv}, a qualitative connection can be made with our study where, in similar manner, the role of quantum phonons is to absorb the excess energy.
For instance, when the coupling to the heat reservoir of a driven Hubbard model~\cite{amaricci} is increased, a shift of the maximal current in the $\bar{j}$--$F$ characteristics is observed accompanied with the increase of $\bar{j}$ in the NDR regime. 
These features are consistent with the influence of phonons in Fig.~\ref{fig3}.
Therefore, this work represents an example how to stabilize a QS state of a driven system where the work done by the field is absorbed exclusively by quantum degrees of freedom within the model.

Our investigations of carrier dynamics under the static electric field are limited to modeling driving-induced
low-energy intraband excitations. Focusing on the time-independent driving does not require consideration of  interband transitions as long as the field, used in our calculations, remains below the threshold value for the dielectric breakdown of a Mott insulator. 
To mimic the situation realized in ultrafast experiments where high-frequency pulses are used, one has to combine
mechanisms emerging due to both intraband as well as interband transitions.  
Still, a  purely quantum mechanical description of  nonequilibrium dynamics in correlated multiband models coupled to phonons remains an outstanding theoretical challenge in the field  of driven  systems.

\acknowledgements

L.V. and J.B. acknowledge stimulating discussions with M. Mierzejewski, P. Prelov\v sek and V. V. Kabanov.
This work has been support by the Program P1-0044 of the Slovenian Research Agency (ARRS) 
and REIMEI project, JAEA, Japan.

\end{document}